\documentclass[runningheads]{llncs}
\usepackage[T1]{fontenc}
\usepackage{graphicx}
\usepackage{booktabs}
\usepackage{tabularx}
\usepackage{array}
\usepackage[colorlinks=true, linkcolor=blue, citecolor=blue, urlcolor=blue]{hyperref}
\usepackage[table]{xcolor}
\usepackage{subfigure}
\usepackage{pgfplots}
\pgfplotsset{compat=1.18}
\usepackage{xcolor}
\usepackage{url}
\usepackage{orcidlink}

\newcommand{\err}[1]{\textcolor{blue!70}{#1}}

\newcommand{\corr}[1]{\textcolor{teal!70}{#1}}
\begin{document}
\title{Simulating Novice Students Using Machine Unlearning and Relearning in Large Language Models}

\titlerunning{Simulating Novice Students Using Machine Unlearning and Relearning}
%
\author{Jiajia Song\inst{1}\orcidlink{0009-0007-7840-5576} \and
Zhihan Guo\inst{1}\orcidlink{0009-0004-7638-6779} \and
Jionghao Lin\inst{1\thanks{Corresponding author.}, 2, 3}\orcidlink{0000-0003-3320-3907}}
\authorrunning{Song et al.}
\institute{
    The University of Hong Kong, Pokfulam Rd, Hong Kong, China\\
    \email{kathrynsjia@gmail.com} \\ \email{zhihang330@connect.hku.hk} \\ \email{jionghao@hku.hk} \and   Carnegie Mellon University, Pittsburgh PA 15213, USA  \and     Monash University, Clayton VIC 3800, Australia 
}
\maketitle              
%
\begin{abstract}
Student simulation can support learning-by-teaching pedagogy where human students (as tutors) teach AI-simulated novice students (as tutees). Recent research often relies on prompt engineering with large language models (LLMs) to simulate novice student behaviour, but it is difficult to keep the AI-simulated student at a stable novice knowledge level. A key reason is that many LLMs are trained to be broadly capable, so even when prompted to ``act like a novice,'' the LLMs can still produce expert-level explanations during the learning-by-teaching interaction process. As a result, the AI-simulated student may drift beyond the intended knowledge level, reducing the credibility of the simulation for studying learning-by-teaching processes. Thus, we propose a knowledge-level simulation approach based on machine unlearning. We investigate this approach using a dataset of multiple-choice questions on Python programming concepts. We apply machine unlearning to transform a knowledgeable LLM into a novice-level AI student (i.e., teachable agent), then evaluate whether the teachable agent can relearn targeted knowledge components through learning-by-teaching dialogue interactions. 
Finally, we analyse the dialogue logs to characterise how the agent’s behaviour changes over time, including its question asking, error patterns, and responsiveness to instruction. The results show that (1) unlearning produces simulated student agents with more novice-like responses than prompt-only baselines, (2) the agents recover a measurable portion of the unlearned knowledge under structured exposure, and (3) dialogue analyses reveal identifiable trajectories of conceptual change and teaching moves that predict learning recovery. Our code is available on GitHub: \url{https://github.com/GEMLab-HKU/Unlearn_and_Relearn}


\end{abstract}

\section{Introduction}



Human artificial intelligence (HAI) collaboration has rapidly become a common interaction paradigm in learning and problem-solving contexts. In educational settings, students increasingly rely on large language models (LLMs) to answer questions, explain concepts, and complete programming tasks. Although such systems provide immediate and fluent responses, growing evidence suggests that direct question–answer (QA) interactions may unintentionally undermine learners' critical thinking and self-explanation abilities \cite{lee2025impact}. When students pose questions, LLMs often respond with confident and affirmative answers \cite{sharma2023towards}, effectively validating learners' initial assumptions rather than requesting self-explanantion, showing misconceptions, or encouraging reflection, which are widely recognised as essential mechanisms for knowledge construction and conceptual change \cite{zhang2024effects}. Research has long shown that effective learning emerges when learners actively construct knowledge through explanation, error detection, and reflection, rather than passively receiving answers \cite{kozanitis2023effect,zhang2025unravelling}. One well-established approach that supports these processes is learning by teaching, in which human students (as tutors) teaching and explaining concepts to others (as tutees) promotes metacognitive monitoring and facilitates conceptual change \cite{zhu2024effects}. Recent advances in LLMs have inspired the use of AI agents to simulate learning by teaching interactions, such as pairing an AI ``teacher'' with an AI ``student'' \cite{rogers2025playing,chen2024learning}. 
These systems aim to encourage explanation-oriented interaction and to harness the benefits of learning by teaching, which has been shown to promote deeper learning, metacognitive awareness, and sustained engagement, and to address the limitations of AI systems that primarily provide direct answers.

A key challenge in developing LLM-based teachable agents for learning-by-teaching interactions is maintaining a stable novice knowledge level. Many existing approaches rely on prompt engineering to guide LLMs to behave as beginners or low-knowledge students \cite{Lu_2024,martynova2025can}. Although such prompt-based simulations of novice students can be effective initially, they often suffer from instability and reproducibility in different settings, as indicated by the prior works \cite{benedetto2024using,rogers2025playing}. We argue that for LLM-based agents to be genuinely teachable, they require model-level knowledge degradation because prompt-based role assignment does not alter the model's underlying pre-trained knowledge, often leading to expert knowledge leakage and unstable role fidelity. 
This loss of role fidelity undermines the effectiveness of learning by teaching and 
limits the analysis of learning behaviors.

To address this limitation, we propose a knowledge-level student simulation approach based on machine unlearning, which aims to remove or suppress specific data or knowledge from the trained model \cite{thaker2025positionllmunlearningbenchmarks}. 
We construct a teachable agent by selectively removing a subset of questions associated with predefined knowledge components, which are the units of forgotten programming knowledge for the teachable agent, drawn from high-frequency concepts in the dataset. As a result, the agent exhibits stable, novice-like behavior and can meaningfully participate in peer-teaching interactions. This 
approach enables controlled investigation of how artificially induced novice agents respond to instruction and whether they can recover forgotten knowledge through structured exposure and interactive teaching, allowing agent learning progress to serve as meaningful evidence of instructional effectiveness and student understanding. The research questions (\textbf{RQ}s) are as follows: 
\begin{itemize}
  \item \textbf{RQ1}: Can machine unlearning reliably induce controllable and stable novice knowledge states in LLM-based teachable agents?
  \item \textbf{RQ2}: To what extent can teachable agents relearn the removed knowledge through structured exposure and interactive teaching?
\end{itemize}

\section{Related Work}
\vspace{-2mm}
\subsection{Learning by Teaching Approach}
\vspace{-1mm}

Learning by teaching a student peer can foster deep and active learning \cite{kobayashi2019learning}. Peer learning is common in higher education, and the benefits become stronger when students adopt the teacher role because they must reorganize knowledge, generate explanations and analogies, and regulate their thinking, which supports metacognition and self-regulation \cite{fiorella2015learning,kobayashi2019learning}. Teaching an audience, even an artificial one, can further sustain reflection as students monitor their understanding, notice gaps, and refine their mental models \cite{yingbin2021can}. Consistent evidence suggests that students exert more effort and achieve greater learning gains when they believe they are teaching a peer or a teachable agent \cite{kobayashi2019learning}.

Despite these advantages, bringing learning by teaching into real classrooms is not straightforward. Without guidance, students may simply give answers instead of prompting reasoning, and they may struggle to tailor explanations to a novice's changing needs \cite{rogers2025playing}. These challenges highlight the need for scalable support to sustain effective learning-by-teaching activities. Student simulation provides one form by offering a controllable novice learner for practice and evaluation, enabling learning-by-teaching interactions at scale.


\vspace{-3mm}
\subsection{Student Simulation}
\vspace{-1mm}
Student simulation refers to computational methods that generate plausible actions or responses of the student while the student works on a task \cite{tian2025large}. In AI in education, student simulators are often used to test instructional approaches and study how different types of support might influence learning processes, without relying on large-scale classroom deployment at the earliest stage \cite{scarlatos2026simulated}. Classical approaches typically encode a student state (e.g., what the learner currently knows, misunderstands, or is likely to do next) and then sample the next step or response conditioned on the task context. A well-known line of work is \textit{SimStudent} \cite{matsuda2015teaching}, a teachable agent that learns cognitive skills from instruction and examples by human students (as tutors) and can then be used to study tutoring strategies and learning-by-teaching settings \cite{matsuda2015teaching}. 
\vspace{-3mm}

\subsection{Teachable Agents}
\vspace{-1mm}
 Teachable agents operationalise the learning-by-teaching approach by giving human students a ``tutee'' (i.e., AI-simulated novice student) to instruct, creating conditions for the protégé effect, where learners invest more effort because they feel responsible for helping the agent improve \cite{chase2009teachable}.
Early teachable-agent systems, including well-known classroom-facing platforms (\textit{Betty's Brain} \cite{biswas2016design} and \textit{SimStudent} \cite{matsuda2022teachable}), showed that students can learn by constructing explanations for an agent, debugging misconceptions, and revising their teaching strategies across interactions \cite{biswas2016design,yingbin2021can,matsuda2022teachable}. 
However, these systems typically rely on hand-crafted domain models and predefined knowledge representations, with interaction grounded in structured tasks or constrained teaching actions. While effective within well-defined domains, such designs limit the systems' ability to engage in open-ended dialogue or scale flexibly to more diverse subject matter.

LLMs introduce new opportunities for teachable-agent design. For example, LLMs can sustain fluent dialogue and can display a wide range of human-like behaviours \cite{jin2024teach}. Recent work has begun to explore LLM-based teachable agents in programming contexts \cite{chen2024learning,ma2024teach}. In contrast to earlier systems that rely on hand-crafted domain models and constrained interaction formats, LLM-based teachable agents can engage in natural language explanations without extensive domain engineering, support richer and more adaptive interaction patterns, and flexibly reflect different levels of knowledge or misconceptions. 
Yet, this shift raises a core concern for learning-by-teaching: if the ``tutee'' remains too competent due to the language model's pre-training, learners may stop explaining, stop checking, and instead accept answers, which undermines the intended benefits of the novice tutee effect. Recent student-simulation work explicitly flags this novice-to-expert drift as a validity threat in LLM-based student simulation, especially when the system relies primarily on prompting rather than changing the model's knowledge state \cite{rogers2025playing,chen2024learning}. 


\vspace{-3mm}
\subsection{Machine Unlearning and Relearning for Teachable Agents}
\vspace{-1mm}
Machine unlearning refers to techniques that remove the influence of specific training data or targeted knowledge from a trained model, aiming to approximate the model that would have been obtained if the removed data had never been used for training \cite{bourtoule2021machine,liu2025rethinking}. This is motivated by data deletion requirements and by safety goals such as reducing unwanted memorisation or capabilities, while avoiding full retraining from scratch \cite{wang2025reasoning}. In the context of student simulation, machine unlearning offers a principled way to address the key limitation of prompt-only simulated students, namely that the underlying competence remains intact. Instead of merely asking a model to ``act like a novice'', unlearning can be used to reduce task-specific competence so the agent's behaviour is more consistently aligned with a novice knowledge level.

An additional, education-relevant question is whether a model that has unlearned task knowledge can later reacquire that knowledge through interaction \cite{xu2025relearn}. This matters because learning-by-teaching is most effective when the tutee's improvement is assessable and contingent on the tutor's effort. Recent evaluation work for LLMs unlearning emphasises multi-faceted criteria for assessing whether forgetting is effective and whether residual traces remain, which connects naturally to studying ``relearning'' dynamics after unlearning in an educational setting. However, the idea of unlearning to initialise a stable novice state and relearning to model growth under tutoring remains underexplored for student simulation in learning-by-teaching contexts.

\section{Methods}

\subsection{Dataset} 
Our study uses a dataset of Python multiple-choice questions (MCQs) collected from publicly available web resources and LLM-generated items for unlearning-oriented training and evaluation. This dataset covers MCQs across a broad set of Python programming concepts such as \textit{Functions}, \textit{Operators}, \textit{Data Types}, and \textit{Lists}. Each question contains a question description, multiple answer options, the correct answer, a concept label, and an explanation. The dataset contains 2,074 MCQs spanning 22 concept categories, combining web-collected questions with LLM-generated items to broaden coverage and increase diversity. 

For the web-collected portion, we crawled publicly available Python MCQs and solutions from established programming resources, including Sanfoundry,\footnote{\url{https://www.sanfoundry.com/}} GeeksforGeeks,\footnote{\url{https://www.geeksforgeeks.org/}}, and PyNative\footnote{\url{https://pynative.com/}} firstly. We also retain metadata to support traceability, which enables auditing the data source. We performed basic cleaning to remove duplicates and to standardize formatting across sources. However, coverage across concepts in these sources is uneven, and some categories include repetitive question styles or limited variation in phrasing. To address these limitations and to balance concept coverage, which is important for controlled unlearning experiments, we supplemented the dataset with LLM-generated MCQs. By using \texttt{ChatGPT-5.1}, we generated additional questions for underrepresented concepts and increased linguistic diversity while enforcing a consistent format across generated question items. We further conduct basic validity screening, including checks for option completeness and answer consistency. This helps ensure the generated MCQs are reliable for training and evaluation.


\subsection{System Implementation}



\begin{figure}[h]
    \centering
    \includegraphics[width=\textwidth]{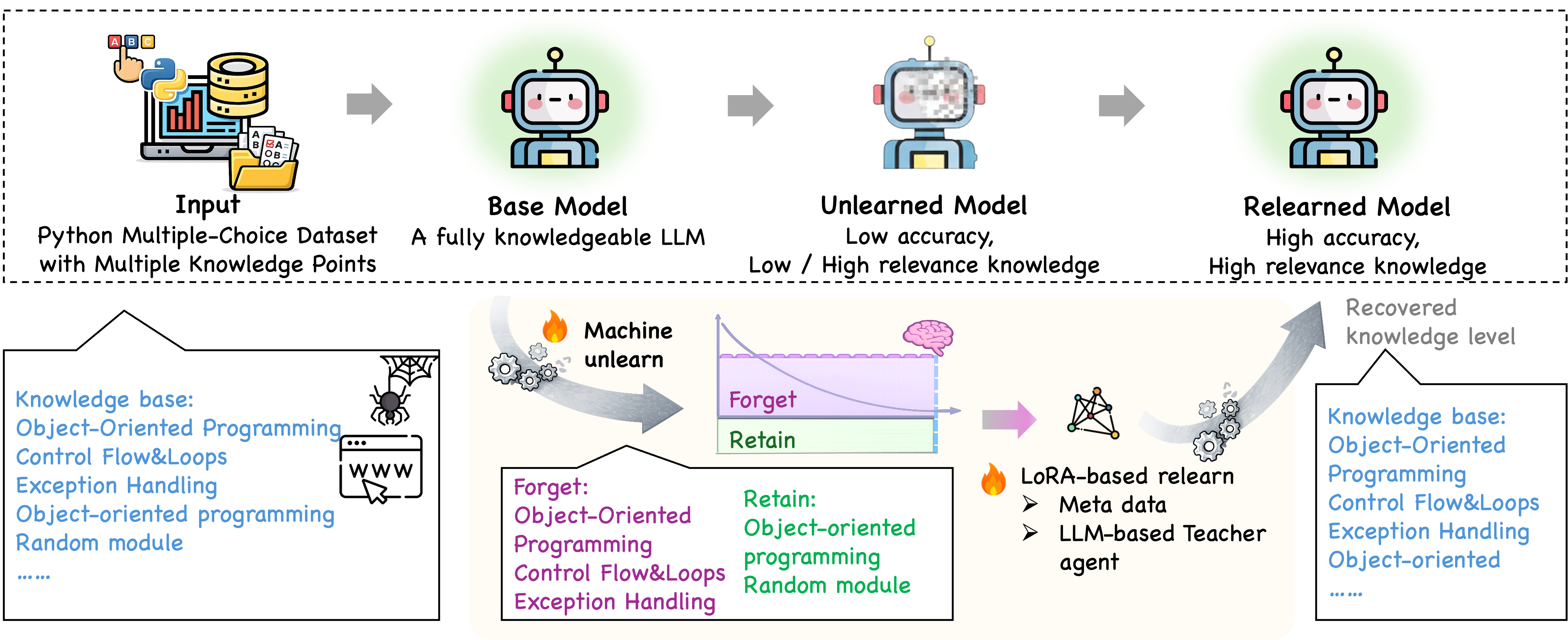}
    \caption{Two-stage training process of the teachable agent's underlying model.}
    \label{fig:model_train}
\end{figure}

Our study is implemented in three stages: \textbf{Stage 1.} Machine Unlearning, \textbf{Stage 2.} Machine Relearning, and \textbf{Stage 3.} Teachable-Agent Interaction. In \textbf{Stage 1} (see Fig.~\ref{fig:model_train}), we adopt an open-source machine unlearning framework \cite{liu2024revisitingwhosharrypotter} and adapt it to our curated Python programming MCQs dataset. 
We propose a two-level data splitting strategy, where the first level operates at the knowledge component (KC) level to determine which concepts are subject to unlearning, and the second level operates at the question level to control the extent of forgetting within the selected KCs. 

\begin{figure}[h]
    \centering
    \includegraphics[width=\textwidth]{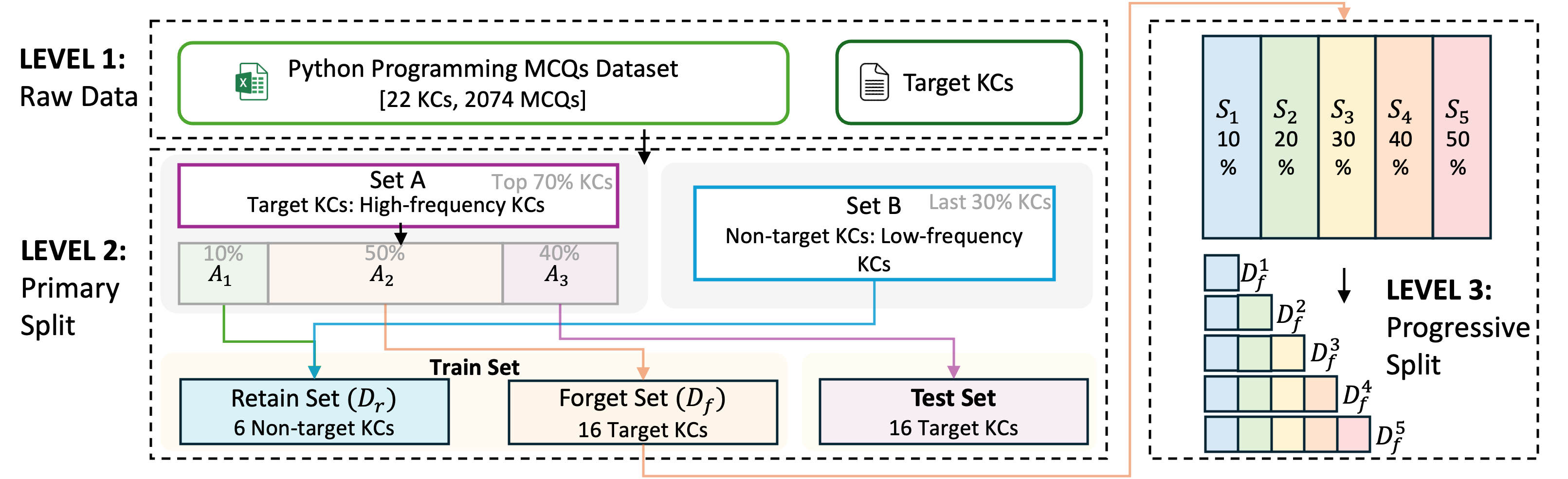}
    \caption{Data preprocessing and splitting strategy for machine unlearning process.}
    \label{fig:data_process}
\end{figure}

As illustrated in Fig.~\ref{fig:data_process}, KCs are first ranked by question count, with the top 70\% (16 KCs, 1,823 questions) forming set A and the remaining 30\% (6 KCs, 251 questions) forming set B. The 70\%/30\% split ensures that unlearning targets are drawn from high-frequency, canonical topics that dominate the dataset, while lower-frequency topics are retained to stabilise overall language and task performance. Questions in set A are then sequentially split into $A_1$ (10\%), $A_2$ (50\%), and $A_3$ (40\%) to balance minimal knowledge retention, effective unlearning, and robust evaluation of forgetting and relearning. The retain set consists of $A_1$ and $B$ (441 samples), the forget set is $A_2$ (903 samples), and the test set is $A_3$ (730 samples). The 16 KCs in set A are treated as target knowledge for unlearning, and $A_2$ can be progressively accumulated (10\%–50\%) to simulate different unlearning levels.
During the unlearning stage, we follow the original optimization objective of the adopted machine unlearning framework \cite{liu2024revisitingwhosharrypotter} and online parameter-efficient fine-tuning (LoRA) the base language model to attenuate representations associated with the designated forgetting set (target-knowledge questions), while maintaining its performance on the retained data (all remaining non-target training samples). LoRA enables effective knowledge attenuation while substantially reducing computational cost compared to full model fine-tuning. Concretely, the model parameters are updated using the forgetting subset as supervision, such that task-relevant knowledge corresponding to the target KC is selectively suppressed rather than globally erased.

Following the machine unlearning (\textbf{Stage 1}), we introduce a controlled relearning stage (\textbf{Stage 2}) to simulate knowledge recovery in an instructional setting. Relearning is implemented in two complementary ways, enabling us to systematically trace how knowledge is recovered under direct supervision versus guided instructional support.
One approach applies standard supervised fine-tuning (SFT) to the unlearned model using the original QA pairs from the forgotten set as supervision, without exposure to any retained samples. 
The cross-entropy loss is computed only over the answer tokens, such that the relearning signal is strictly confined to recovering the suppressed answer content rather than adapting to the question prompt. As a result, improvements in task accuracy reflect the re-acquisition of the target knowledge itself, rather than superficial gains from question-format familiarity or prompt-level adaptation.
This design more closely reflects real educational assessment practices, where learners are evaluated based on their responses rather than their interpretation of the question wording, thereby enabling a more focused analysis of conceptual relearning. Notably, the amount of data used for relearning is progressively aligned with the degree of prior unlearning (e.g., a 10\% unlearning setting is followed by relearning on the corresponding 10\% subset). Through this controlled setup, we are able to evaluate both the thoroughness of unlearning—by examining whether forgotten knowledge can be recovered—and the model’s capacity to relearn previously removed knowledge under subsequent training.

\begin{figure}[t]
    \centering
    \includegraphics[width=\textwidth]{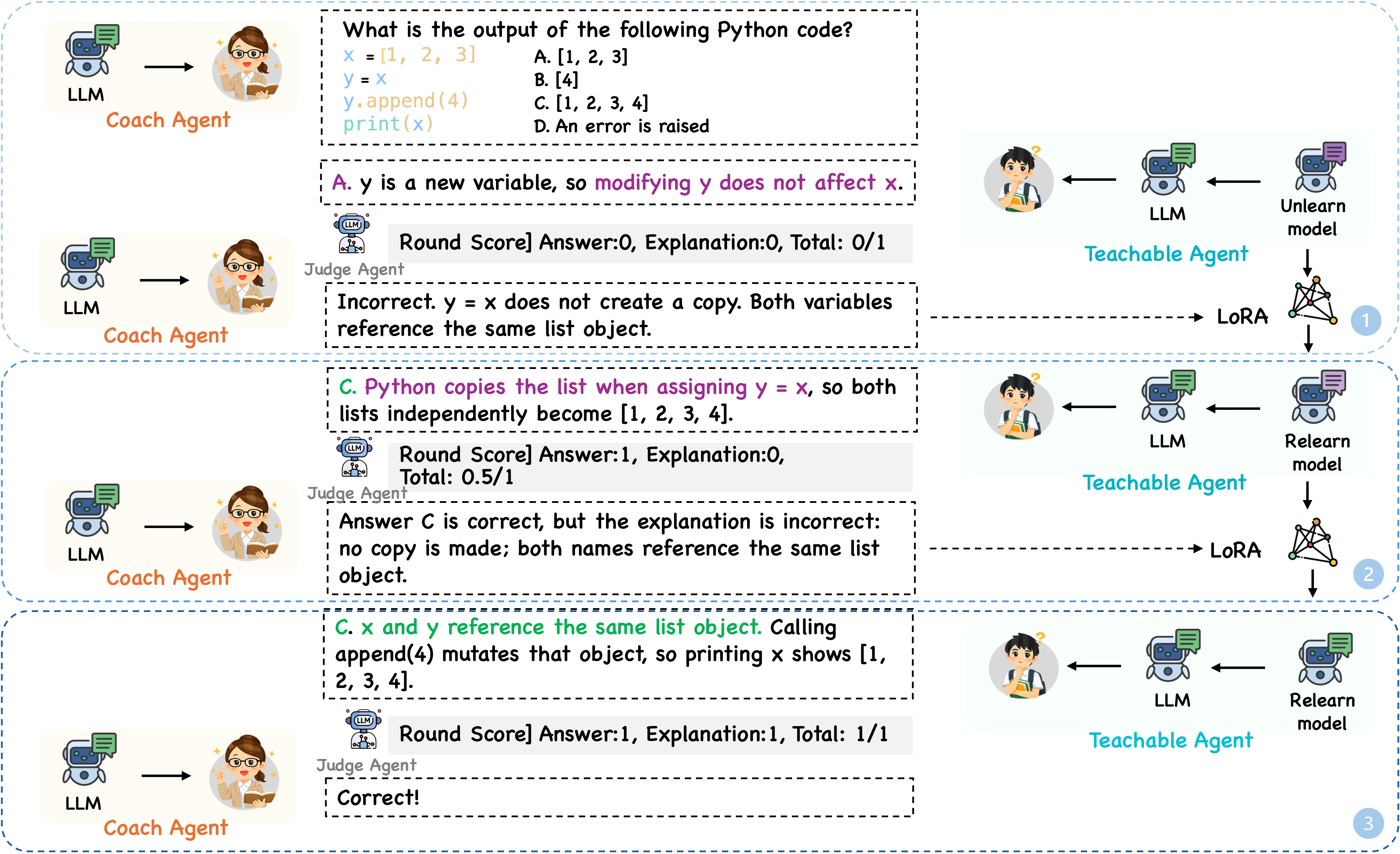}
    \caption{Simulating a three-round interaction process between a coach agent and a teachable agent.}
    \label{fig:agent_simulate}
    \vspace{-3mm}
\end{figure}

Additionally, in \textbf{Stage 3}, we adopt an instructional relearning mechanism based on an LLM-based coach agent within a teachable-agent interaction setting. In this setting, the coach agent engages in dialogue interaction with the teachable agent, providing corrective feedback and conceptual explanations in response to the teachable agent’s incorrect answers. These coach-generated explanations are then used as training signals to further fine-tune the unlearned model, simulating knowledge recovery through guided instruction rather than direct exposure to ground-truth answers.
Specifically, as illustrated in Fig.~\ref{fig:agent_simulate}, the unlearned model is deployed as a teachable agent within a three-agent interactive learning environment, designed to simulate learning by teaching under asymmetric knowledge conditions. The teachable agent is instantiated from a knowledge-suppressed (unlearned) model while the coach agent retains full task knowledge from an LLM, thereby ensuring a genuinely novice student state and enabling controlled examination of instructional effects.
The teachable agent, instantiated from the unlearned model, represents a low-knowledge student, initially exhibiting low answer accuracy and low conceptual relevance. It attempts to answer programming MCQs and provide explanations, which are evaluated by a \textit{judge agent} that jointly assesses answer correctness and explanation quality. Based on this evaluation, a \textit{coach agent} powered by \texttt{DeepSeek-V3.2} \cite{liu2025deepseek}  delivers targeted instructional feedback to correct misconceptions. For interaction realism, the student's raw responses are rewritten by the \texttt{DeepSeek-V3.2} to improve grammar and tone, while maximally preserving the underlying semantic content. If full mastery is not achieved, the dialogue history and instructional feedback are used to perform LoRA on the student model, after which the student re-attempts the same question. Through this iterative cycle of response, evaluation, instruction, and update, the student agent progressively transitions from low-accuracy, low-relevance responses to high-accuracy, high-relevance explanations, demonstrating successful acquisition of the target knowledge point through peer-guided learning.

\subsection{Study Setup}
All unlearning experiments are conducted on \texttt{Mistral-7B-Instruct-v0.3} \cite{chaplot2023albert}. 
Unlearning is performed using an intervention-based forgetting loss guided by a teacher distribution, optimized with KL divergence ($\beta = 0.1$) and a retention strength of $1.0$. 
Training is carried out for 20 epochs with a learning rate of $1\mathrm{e}{-4}$ and batch size $8$, using LoRA fine-tuning ($r=8$, $\alpha=32$). 
The teacher distribution is constructed by replacing each correct answer with three plausible incorrect alternatives ($N=3$), without counterfactual prompting.


\vspace{-2mm}
\section{Results}
\vspace{-1mm}
\subsection{Machine unlearning enables novice-level student simulation}

\begin{figure}[t]
    \centering
    \subfigure[]{
        \includegraphics[width=0.475\textwidth]{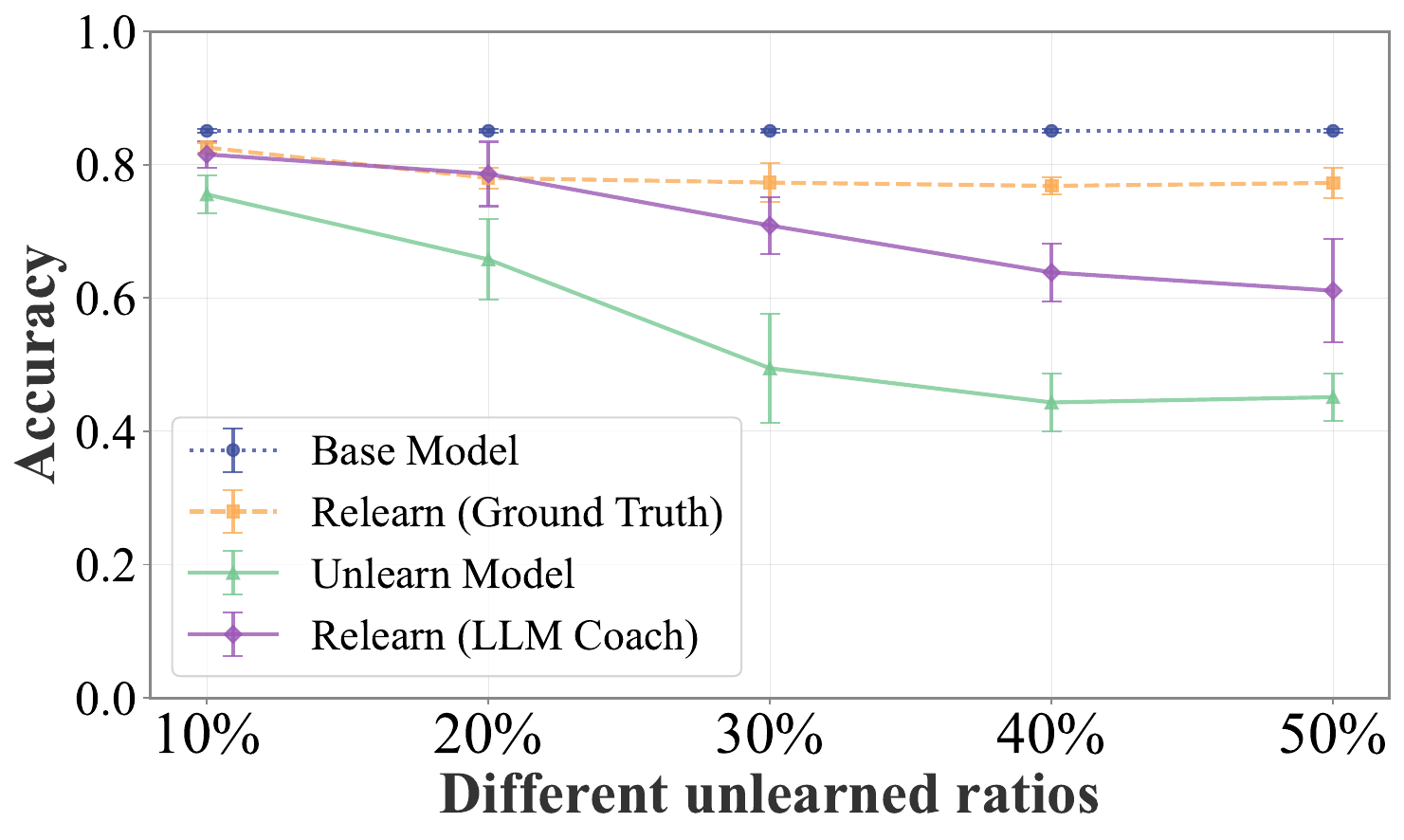}
        \label{fig:agent_simulate_a}
    }
    \subfigure[]{
        \includegraphics[width=0.475\textwidth]{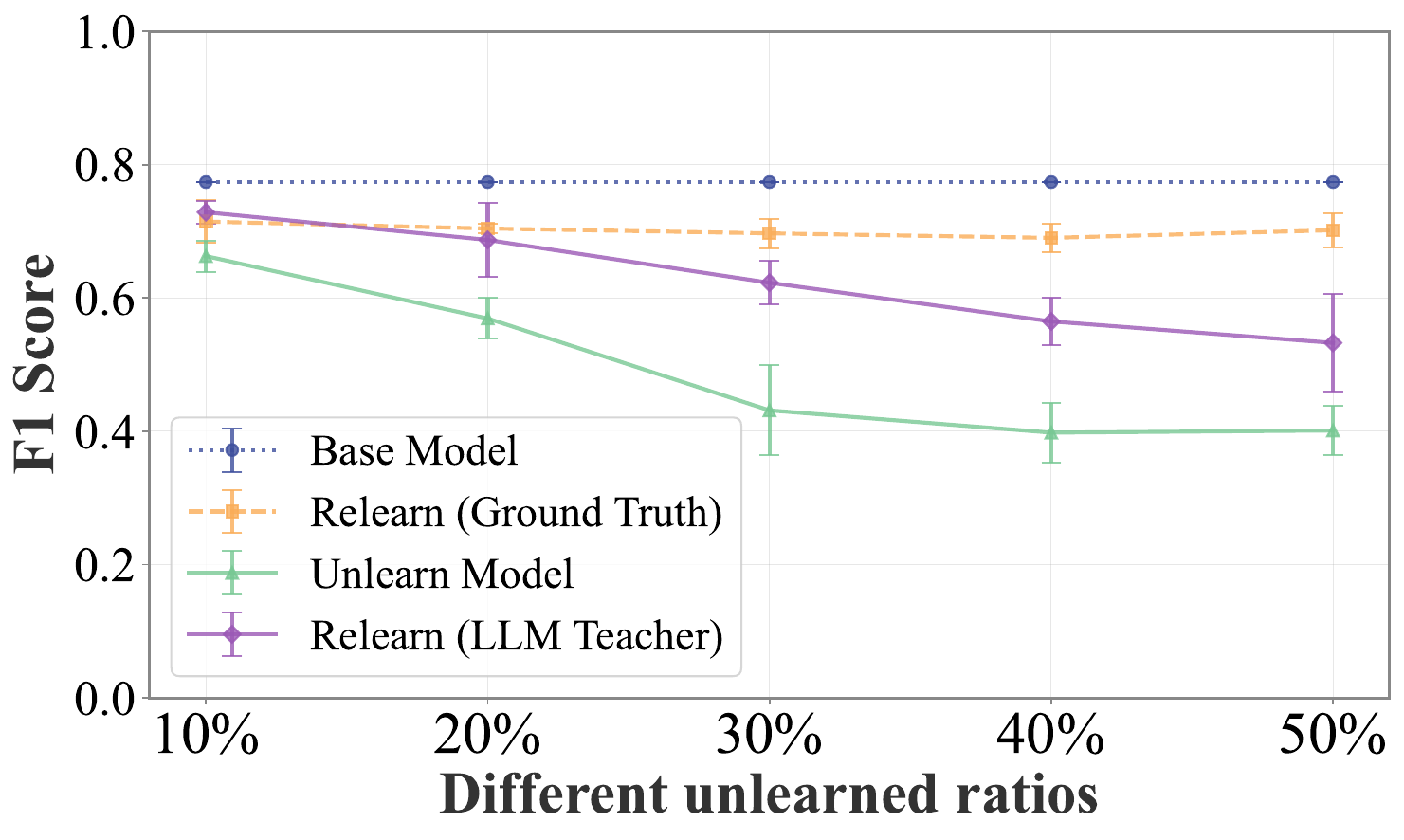}
        \label{fig:agent_simulate_b}
    }
    \caption{
Performance under unlearning ratios from 10\% to 50\%: (a) classification accuracy on MCQ answering and (b) F1 score. Error bars indicate the standard deviation across five random seeds.}
    \label{fig:results}
    \vspace{-4mm}
\end{figure}

To answer RQ1, we evaluate the forgetting capability of the model under five random seeds to obtain more robust and reliable results. 
As shown in Fig.~\ref{fig:results}, we compare both answer accuracy and F1 score of the base model and the unlearned model under different unlearning ratios (10\%–50\%).
The unlearned model (green line) exhibits a clear and consistent decline in both accuracy and F1 score as the unlearning ratio increases, dropping from around 0.75 at 10\% to below 0.5 from the 40\% unlearning ratio onward. This pattern suggests that the drop in accuracy is caused by the model progressively losing the targeted knowledge as unlearning becomes stronger, rather than by random variation or evaluation noise.
In contrast, the base model (blue line), defined as the original pretrained model without any further fine-tuning or modification for a specific task, maintains a consistently high accuracy (around 0.85) and F1 scores (around 0.77) across all unlearning ratios. This stability indicates that the observed performance degradation is not due to randomness or evaluation variance, but is directly attributable to the unlearning procedure.

\vspace{-3mm}

\subsection{Teachable agents relearn through learning-by-teaching}
\vspace{-2mm}

As shown in Fig.~\ref{fig:agent_simulate_a}, unlearning leads to a clear and monotonic degradation in answer accuracy as the forgetting ratio increases from 10\% to 50\%, confirming that the unlearning procedure effectively suppresses target knowledge. However, this degradation is not irreversible. Both relearning strategies consistently improve performance across all unlearning ratios. Specifically, relearned models (orange line) substantially outperform their corresponding unlearned counterparts, demonstrating that previously suppressed knowledge can be reacquired rather than being completely erased. Nevertheless, the relearned models do not fully recover to the performance level of the base model (blue line), and the performance gap widens under higher unlearning ratios (30\%–50\%), indicating that knowledge recovery is measurable yet incomplete, especially when a larger portion of knowledge has been unlearned. Relearn(LLM Coach) highlighted in purple line in Fig.~\ref{fig:results} further examines whether interactive teaching can facilitate knowledge recovery in a more realistic instructional setting. The unlearned model is deployed as a teachable agent and guided by an LLM-based coach through multi-round dialogue. Compared with static relearning, coach-guided interaction yields consistent and progressive gains in accuracy across unlearning ratios, particularly under moderate forgetting levels (10\%–30\%). This suggests that instructional feedback helps the learner agent reconstruct missing conceptual links rather than merely memorizing answers.

Moreover, the F1 score captures both precision and recall and is therefore more sensitive than accuracy to imbalanced or inconsistent predictions induced by unlearning. As shown in Fig.~\ref{fig:agent_simulate_b}, the unlearned model exhibits a clear and monotonic decline in F1 score as the unlearning ratio increases, indicating not only reduced correctness but also degraded stability in answer selection. Both relearning strategies consistently improve F1 scores across all ratios, confirming partial recovery of suppressed knowledge. However, compared with accuracy, the performance gap to the base model remains more pronounced in F1, especially under higher unlearning ratios (30\%–50\%), suggesting that deeper conceptual inconsistencies persist. Relearn (LLM Coach) achieves consistently higher F1 scores than static relearning, highlighting the advantage of interactive teaching in promoting more balanced and coherent knowledge recovery.

\begin{figure}[h]
    \centering
    \includegraphics[width=0.9\textwidth]{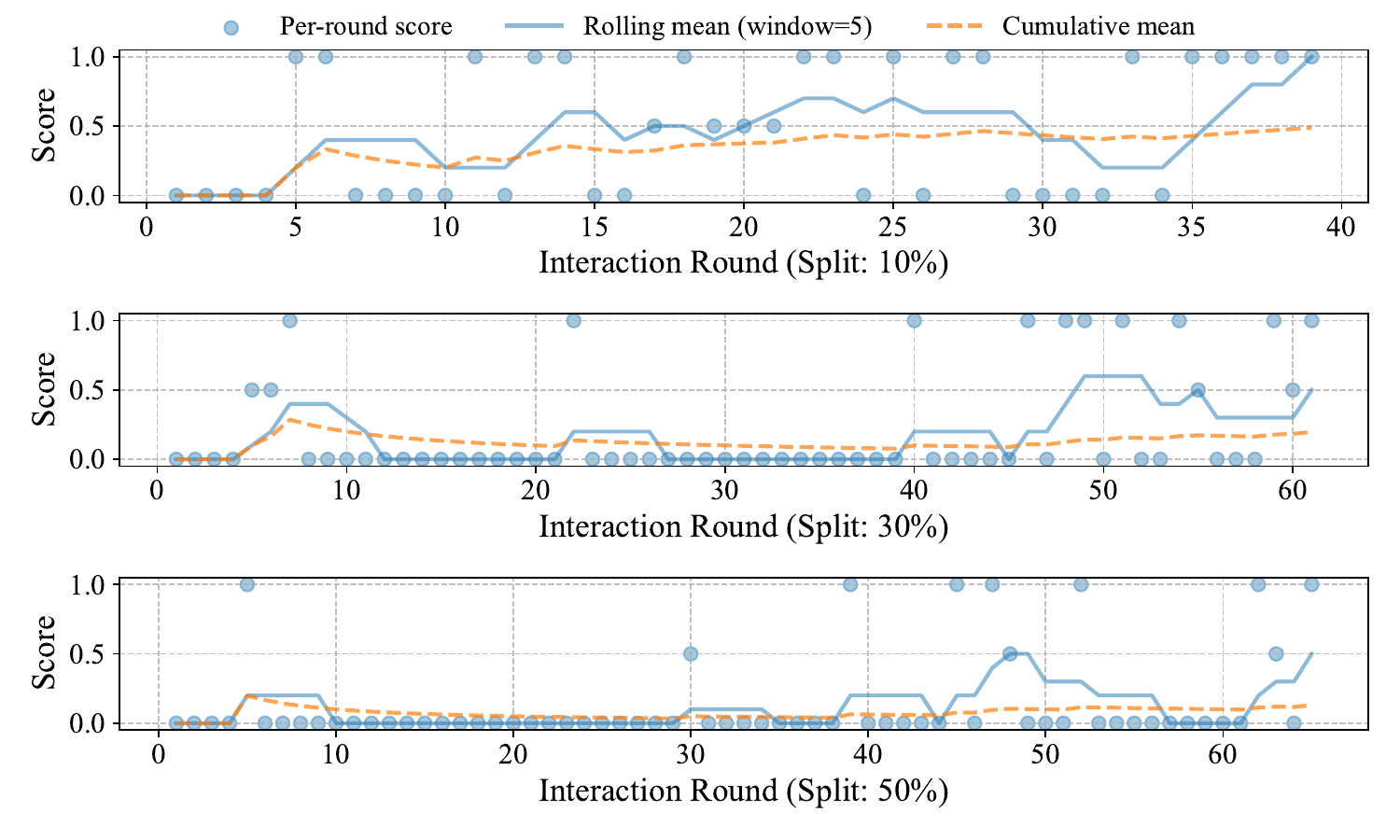}
    \caption{Comparison of answer correctness scores (combining answer selection and explanation correctness) among models with different unlearning ratios (10\%, 30\%, and 50\%) under instruction from an LLM-based coach agent on the topic of Exception Handling. These ratios are selected as representative settings for clarity of presentation; results for other unlearning ratios can be found in the  \href{https://github.com/GEMLab-HKU/Unlearn_and_Relearn/blob/main/pic/pic3.png}{GitHub Repo}.}
    \label{fig:split_compare}
    \vspace{-3mm}
\end{figure}

We further conducted qualitative analysis of how relearning unfolds through interaction. 
In Fig.~\ref{fig:split_compare}, teachable agents with a lower unlearning ratio (10\%) begin to exhibit performance recovery within the early interaction rounds, followed by a steady increase in both rolling and cumulative accuracy. In contrast, agents with higher unlearning ratios (30\% and 50\%) display prolonged low-accuracy plateaus, sporadic correct responses, and delayed improvement, indicating weaker retention of executable conceptual structure and less stable relearning dynamics. 
\vspace{-5mm}

The dialogue-level evidence in Table~\ref{tab:dialogue_comparison_exception} further clarifies the nature of this improvement. Before instruction, the learner provides a fluent but conceptually incorrect explanation of exception chaining, misinterpreting the \textbf{\texttt{from}} keyword as a security mechanism for hiding errors. After targeted feedback from the coach agent, the learner not only selects the correct option but also provides a conceptually accurate explanation of the underlying mechanism. This shift reflects a correction of the underlying mental model rather than superficial answer alignment, demonstrating genuine conceptual change induced through interaction.

\begin{table}[h]
\centering
\caption{Dialogue comparison of student understanding before and after teaching (10\% unlearning split).}
\label{tab:dialogue_comparison_exception}
\renewcommand{\arraystretch}{1.7}
\scriptsize
\begin{tabular}{@{}p{2.4cm}@{\hspace{0.25cm}}p{4.3cm}@{\hspace{0.25cm}}p{4.3cm}@{}}
\hline
\rowcolor{gray!10}
\multicolumn{3}{@{}p{12cm}@{}}{%
\textbf{Anchor question:}
\textit{What is the main purpose of writing
\texttt{raise NewError(`msg') from original\_error}
in a Python \texttt{except} block?}

\par A.~To completely hide \texttt{original\_error} from the traceback so that only \texttt{NewError} is visible to the end user, improving system security by design.

\par B.~To explicitly link \texttt{NewError} to \texttt{original\_error} so that the traceback shows both the higher-level context and the underlying cause, which helps when debugging layered or wrapped code.

\par C.~To force the interpreter to retry the failing operation using the type of \texttt{NewError} instead of \texttt{original\_error}, effectively implementing an automatic recovery mechanism.

\par D.~To convert \texttt{original\_error} into a warning while still raising \texttt{NewError}, causing logging frameworks to treat the situation as non-fatal by default.
} \\
\hline
\textbf{Dimension} & \textbf{Before Teaching} & \textbf{After Teaching} \\
\hline
Selected Answer &
Option A &
Option B \\
\hline
Original Explanation &
\textit{
\err{Hiding the error message} helps prevent
\err{unauthorized users or attackers} from accessing
\err{sensitive information in the stack trace},
which improves overall \err{system security}.
} &
\textit{
Using \texttt{... from ...} \corr{preserves and links}
the original error, showing both the higher-level error
and the \corr{underlying cause}, which aids debugging.
} \\

\hline
Core Idea &
\texttt{from} is used to hide the original error for security purposes &
\texttt{from} is used to explicitly preserve and link the exception cause \\
\hline
Correctness &
Incorrect &
Correct \\
\hline
Error Cause &
Misunderstanding of exception chaining semantics &
-- \\
\hline
\end{tabular}
\vspace{-5mm}
\end{table}

\vspace{-2mm}

\section{Discussion}
\vspace{-2mm}
Our results show that machine unlearning can be used to adjust the LLM-based teachable agent's knowledge level.
As more knowledge is removed, the agent becomes less accurate and its explanations become less conceptually clear.
Qualitative analyses further reveal a progression from partial misconceptions to more persistent misunderstandings, indicating genuine conceptual disruption rather than random noise. Importantly, this degradation is not irreversible.
The agents recovered part of the suppressed knowledge under both static relearning, where the model was retrained on the original forgotten QA pairs, and coach-guided interaction, where an LLM-based coach supported the agent through multi-round dialogue with corrective feedback and conceptual explanations. Compared with static relearning, dialogue-based instruction produced more stable and progressive gains, especially under moderate unlearning.
These findings suggest that LLM-based student simulation provides a feasible and interpretable framework for studying instructional effects on conceptual change.

\subsection{Implications}


\textbf{Stable LLM-simulated novice student.} This work demonstrates a way to build novice simulation (or teachable agent) by shaping what the agent actually knows or misunderstands rather than relying only on prompt-level role assignment. When an agent begins from a genuinely limited knowledge state, it is more likely to behave like a novice student who needs explanation, correction, and support. This is important because the teachable agent can then learn in a more meaningful way from the human student, instead of simply producing fluent responses without a real need for instruction. Drawing on the prot\'eg\'e effect \cite{chase2009teachable}, students may also learn more deeply when they take responsibility for helping another learner improve. As they observe the agent making progress in response to their explanations, they may become more willing to invest effort, elaborate their reasoning, and provide more thoughtful instruction. This can create a positive feedback loop: stronger explanations from the student lead to visible gains in the agent's understanding, and those visible gains may in turn encourage the student to engage more deeply in teaching.

\noindent \textbf{Potential for process assessment.} This work also points to a new possibility for process-based assessment through teachable agents. Instead of evaluating human students solely based on their own task performance, student understanding can also be inferred from how effectively they support a novice agent's learning during the interaction. When a human student helps the agent achieve sustained and meaningful progress, the student can explain concepts, diagnose misunderstanding, and provide feedback in a conversational manner. These interaction traces can therefore be analyzed to assess multiple aspects of student competence, including conceptual understanding, explanation quality, responsiveness to error, and pedagogical decision-making. In this sense, learning can be assessed not only through direct answers or test scores, but also through the quality and consequences of the student's learning-by-teaching process.

\noindent \textbf{Support for professional development.} Beyond student simulation, the findings suggest a practical application for professional development. Because the unlearning approach can generate simulated students with different knowledge levels as reflected in Fig. \ref{fig:results}, it could support training environments in which teachers or tutors practice responding to varied learner needs. This would allow educators to rehearse key pedagogical skills across a wider range of learner profiles than is typically feasible in role-play or live coaching. From this perspective, unlearning-based simulation offers a scalable infrastructure for practice-based professional development.

\subsection{Limitations and Future Works}
Despite the meaningful exploration presented in this study, several limitations remain. First, our work focuses primarily on Python programming concepts and multiple-choice question formats, and all experiments are conducted using a single base language model, which may limit the generalisability of our findings to other subjects, task formats, or model architectures. 
To address this limitation, future work will extend the evaluation to additional domains, such as mathematics, as well as to other task formats, including open-ended questions. It will also compare the proposed unlearning-based simulation with alternative methods, such as prompt-based methods (e.g., zero-shot, few-shot), and evaluate its robustness across different model architectures, such as \texttt{Qwen3-4B-Instruct} \cite{qwen3technicalreport}.
Second, although machine unlearning provides a controllable approach to approximating a novice knowledge state, it primarily operates at the level of knowledge representations and may not capture complex cognitive and social factors involved in human learning, such as affective processes. A necessary next step is to examine the educational applicability of this method through human studies, particularly whether it can support improved learning outcomes and better reflect authentic AI-assisted teaching practices.

\section{Conclusion}
Our work explores machine unlearning as a way to simulate novice students (i.e., teachable agents or novice student agents) in learning-by-teaching scenarios with large language models. By selectively removing targeted programming concepts, we construct student agents whose knowledge states are more stable than those produced through prompt-based role assignment alone. Our quantitative and qualitative findings show that increasing unlearning strength leads to systematic changes in accuracy, explanation quality, and error patterns, yielding behaviors that more closely reflect genuine novice conceptual states. We also show that these agents can recover the unlearned knowledge through both supervised relearning and coach-guided dialogue. Thus, these results suggest that machine unlearning provides a useful foundation for building credible novice student simulations, examining instructional effects in learning-by-teaching, and developing teachable agents with educationally meaningful knowledge states.

\section*{Acknowledgment}

This work was supported by the Faculty Research Fund and by the grant from the URC (Grant No. 2401102970) at The University of Hong Kong.  The opinions, findings, and conclusions expressed in this paper are solely those of the authors.

\bibliographystyle{splncs04}
\bibliography{mybibliography}

%




\end{document}